\documentclass[twocolumn,showpacs]{revtex4}
\usepackage{graphicx}
\usepackage{dcolumn}

\begin{document}


\title{Competition between isoscalar and isovector pairing correlations in $N=Z$
        nuclei}

\author{K. Kaneko$^{1}$ and M. Hasegawa$^{2}$}
\affiliation{
$^{1}$Department of Physics, Kyushu Sangyo University, Fukuoka 813-8503, Japan \\
$^{2}$Laboratory of Physics, Fukuoka Dental College, Fukuoka 814-0193, Japan \\
}

\date{\today}

\begin{abstract}

We study the isoscalar ($T=0$) and isovector ($T=1$) pairing correlations in $N=Z$ nuclei. 
They are estimated from the double difference of binding energies for odd-odd $N=Z$ 
nuclei and the odd-even mass difference for the neighboring odd-mass nuclei, respectively. 
The empirical and  BCS calculations based on a $T=0$ and $T=1$ pairing model reproduce 
well the almost degeneracy of the lowest $T=0$ and $T=1$ states over a wide range of 
even-even and odd-odd $N=Z$ nuclei. It is shown that this degeneracy is attributed to 
competition between the isoscalar and isovector pairing correlations in $N=Z$ nuclei. 
The calculations give an interesting prediction that the odd-odd $N=Z$ nucleus $^{82}$Nb 
has possibly the ground state with $T=0$. 

\end{abstract}


\maketitle

There is a current topic with increasing interests in studying isovector ($T=1$) 
and isoscalar ($T=0$) proton-neutron ($pn$) pairing correlations in $N=Z$ nuclei \cite{Satula1}. 
At present, it is not clear 
whether $pn$ pairing correlations are strong enough to form a static condensate. 
It is well known that an experimental signature of like-nucleon 
proton-proton ($pp$) and neutron-neutron ($nn$) $J=0$ pairing correlations in nuclei with 
neutron excess is the odd-even 
mass difference, which is extra binding energy of even-even nuclei relative to 
that of odd-mass nuclei. However, the odd-even mass differences for even-even $N=Z$ nuclei are 
larger than those of the neighboring even-even $N=Z+2$ nuclei, and it reflects 
the gain in pairing due to stronger $pn$ correlations \cite{kaneko}. 
It has recently been shown \cite{Satula2,Dobaczewski} that the three-point odd-even mass 
difference for an odd-mass nucleus with neutron excess is an excellent measure of $pp$ 
and $nn$ pairing correlations in neighboring even-even nucleus, 
although it is still controversial \cite{Duguet}. 
This conclusion suggests that the $pp$ and $nn$ pairing correlations 
in $N=Z$ even-even nuclei also can be estimated from the odd-even mass 
difference of neighboring odd-mass nuclei with $N=Z+1$. 
On the other hand, the $pn$ pairing can be estimated from the double difference of 
binding energies \cite{kaneko}. 
When we assume isospin symmetry 
in $N\approx Z$ nuclei, 
the $T=1$ $pn$ pairing and like-nucleon ($pp$ and $nn$) 
pairing are classified in the same $T=1$ pairing correlations, and the former correlation energy 
should be the same as the latter one. 

Odd-odd $N=Z$ nuclei are an ideal experimental laboratory for the study of $pn$ pairing 
correlations. 
It is well known that the lowest $T=0$ and $T=1$ states in odd-odd 
$N=Z$ nuclei are almost degenerate and exhibit the inversion of the sign of the energy difference 
$E_{T=1}-E_{T=0}$, 
while all even-even $N=Z$ nuclei have the $T=0$ ground states and the $T=1$ excited states with large 
excitation energies. 
Several authors \cite{Vogel,Janecke1,Zeldes,Macc,Janecke2,Frauendorf} 
already pointed out that this degeneracy in odd-odd $N=Z$ nuclei 
reflects the delicate balance 
between the symmetry energy and the pairing correlations. 
The $T=0$ and $T=1$ ground-state binding energies of $N=Z$ nuclei were calculated by 
using an algebraic model based on IBM-4 \cite{Baldini}. 
In this paper, we study the $T=0$ and $T=1$ pairing correlations from a phenomenological point 
of view, and analyze them in the BCS calculations within a schematic model 
that includes $T=1$ and $T=0$ pairing interactions. 

We begin with the estimation of $T=1$ pairing correlations in $N=Z$ nuclei. 
A typical indicator for $T=1$ pairing correlations is the following three-point odd-even mass 
difference: 
\begin{eqnarray}
\Delta^{(3)}_{n}(Z,N) & = & \frac{(-1)^{N}}{2}[B(Z,N+1) \nonumber \\
                      & - & 2B(Z,N)+B(Z,N-1)], \label{eq:1}
\end{eqnarray}
where $B(Z,N)$ is the negative binding energy of a system. 
Since $B(Z,N\pm 1)\approx B(Z,N)+\Delta \pm \lambda$ based on standard BCS theory 
with pairing gap $\Delta$ leads to $\Delta^{(3)}_{n}(Z,N)\approx \Delta$, 
the indicator $\Delta^{(3)}_{n}$ is often interpreted as a measure of the empirical pairing gap. 
However, it is well known that values of $\Delta^{(3)}_{n}(Z={\rm even},N)$ are large 
for even-$N$ and small for odd-$N$. 
It was discussed \cite{Satula2} that $\Delta^{(3)}_{n}(Z={\rm even},N={\rm odd})$ is an excellent 
measure of $T=1$ pairing correlations, and the differences of $\Delta^{(3)}_{n}$ at 
adjacent even- and odd-N nuclei reflect the mean-field contributions. 
From a view point of the semi-empirical mass formula, the above indicator is well known to be 
affected by the symmetry energy term in the liquid-drop model. In the macroscopic-microscopic 
shell model, however, the curvature contribution cancels out the symmtery 
energy contribution as pointed out by Satulta {\it et al.}\cite{Satula2}. 
What does the magnitude of the pairing gap in the $N=Z$ nuclei mean? 
We suggest that $\Delta^{(3)}_{n}(Z,Z+1)$ of odd-mass nucleus 
should be regarded as pure pairing gap in $N=Z$ adjacent even-even and odd-odd nuclei. 
For the $N=Z$ nuclei, the four and five point indicators cannot be adopted because they include 
large contributions from mean filed and $pn$ correlations \cite{kaneko,Duguet}. 
Figure 1 shows experimental values of $\Delta^{(3)}_{n}$ in odd-mass nuclei, where we plot 
$\Delta^{(3)}_{n}(Z,Z+1)$ for $16<A<60$. 
When there is no data of $\Delta^{(3)}_{n}(Z,Z+1)$ for $60<A<110$, we adopt 
$\Delta^{(3)}_{n}$ for nearest nuclei with $N=Z+1$. 
The expected quenching of neutron pairing at magic (or semi-magic) particle number $N$ or 
$Z$=14, 28, 40, and 50 is clearly seen in the figure. 
\begin{figure}[t]
\includegraphics[width=8cm,height=8cm]{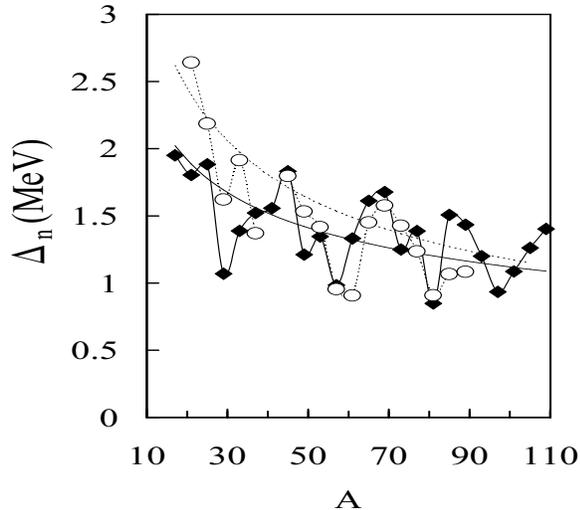}
  \caption{The experimental odd-even mass differences 
  $\Delta^{(3)}_{n}(Z={\rm even},Z+1)$ (solid diamonds) 
  in odd-mass nuclei with $N=Z+1$, and 
  the pairing gaps (open circles) obtained by the BCS 
  calculations. The solid curve is $5.18A^{-1/3}$ and the dashed curve denotes 
  $12A^{-1/2}$.}
  \label{fig1}
\end{figure}
The standard curve $12A^{-1/2}$ is also shown as a guide eye in Fig. 1. 
We can see that the average pairing gap is smaller than the values of the curve $12A^{-1/2}$. 
The global trend can be fitted by the curve $5.18A^{-1/3}$ MeV, 
as discussed in 
recent analyses \cite{Duflo,Vogel}, where $T=1$ pairing gap $\Delta_{T=1}$ obtained 
from some binding energy difference is fitted by the mass-dependence $A^{-1/3}$ different 
from the standard one $12A^{-1/2}$. 
The difference between the two curves is 
quite large for light nuclei, while it is small for heavy nuclei. 
The average gap was recently analyzed \cite{Hilaire} by $\Delta=\alpha+\beta A^{-1/3}$ 
which has theoretical foundation. This analysis also supports the weaker mass-dependence. 
We now consider the following pairing Hamiltonian to describe the $T=1$ pairing correlations: 
\begin{eqnarray}
H   & = & H_{0} + H_{P} 
     =  \sum_{\alpha}\varepsilon_{a}c_{\alpha}^{\dagger}c_{\alpha} 
      - \frac{1}{2}G\sum_{\kappa}P_{\kappa}^{\dagger}P_{\kappa}, \label{eq:2} 
\end{eqnarray}
where $\varepsilon_{a}$ is the single-particle energy and $P_{\kappa}$ is the $J=0$ 
pair operator with isospin $T=1, T_{z}=\kappa$. Implying isospin invariance to the above 
Hamiltonian, the pairing part $H_{P}$ includes the isovector $pn$ interactions. 
The standard BCS calculations with the pairing Hamiltonian (\ref{eq:2})
were performed in $sd$ and $fpg$ shells. 
We adopted single-particle energies from a spherical Woods-Saxon 
potential in the BCS calculations. 
The pairing force strength $G=24.5/A$ was chosen so as 
to fit the experimental odd-even mass difference $\Delta^{(3)}_{n}(Z={\rm even},Z+1)$ in odd-mass 
nuclei. The BCS results for $A > 40$ almost agree with the experimental 
odd-even mass differences, and moreover reproduce the shell effects. 
The BCS calculations reproduce well the behavior of the observed odd-even mass difference over 
a wide range of $N=Z$ nuclei. 
Thus the $T=1$ pairing correlations can be estimated from the odd-even mass difference 
$\Delta^{(3)}_{n}(Z={\rm even},Z+1)$ in odd-mass nuclei. 
\begin{figure}[b]
\includegraphics[width=8cm,height=8cm]{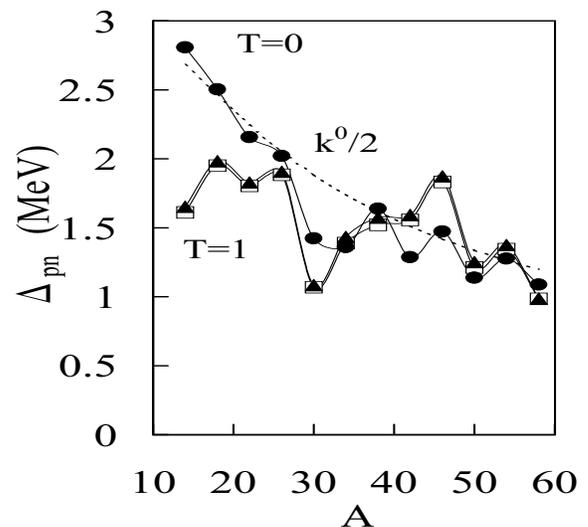}
  \caption{The $pn$ pairing gaps estimated from the double differences of experimental 
  binding energies. The solid circles denote the $T=0$ $pn$ pairing gap, and 
  the solid triangles the the $T=1$ $pn$ pairing gap. The odd-even mass differences 
  in odd-mass nuclei with $N=Z+1$ are shown by the open squares. The dashed curve is 
  the half of the $T=0$ pairing force strength $k^{0}$. }
  \label{fig2}
\end{figure}

To describe the $pn$ pairing correlations in odd-odd $N=Z$ nuclei, let us estimate the 
following double difference of binding energies \cite{Janecke3,Jensen,kaneko}: 
\begin{eqnarray}
\Delta_{pn}^{T}(Z,N) & = & \frac{1}{2}[B(Z,N)^{T} - B(Z,N-1) \nonumber \\
   & - &  B(Z-1,N)+B(Z-1,N-1)], \label{eq:3}
\end{eqnarray}
where $B(Z,N)^{T}$ is the binding energy of lowest state with isospin $T$ in odd-odd 
$N=Z$ nuclei. 
Figure 2 shows the double difference of binding energies calculated from 
the experimental binding energies. The odd-even mass differences 
 for odd-mass nuclei are also displayed. Then we can see that the 
$\Delta^{(3)}_{n}(Z={\rm even},Z+1)$ agrees with the $\Delta_{pn}^{T=1}(Z+1,Z+1)$.  
This means that $T=1$ $pn$ pairing for odd-odd $N=Z$ nuclei have the 
same correlation energy as the like-nucleon $nn$ pairing, $\Delta_{n}=\Delta_{pn}^{T=1}$, 
when assuming isospin symmetry. 
Thus, the indicator $\Delta_{pn}^{T=1}$ gives the $T=1$ $pn$ pairing gap 
in $N=Z$ nuclei. The $\Delta_{pn}^{T=0}$ can be regarded as the $T=0$ $pn$ pairing 
gap as well. 
Figure 2 with these estimations indicates that the $T=0$ $pn$ correlations 
are superior to the $T=1$ $pn$ correlations in the ground states of $sd$ 
shell nuclei, and the inversion occurs in the $pf$ shell nuclei. 
The $T=0$ $pn$ pairing gap $\Delta_{pn}^{T=0}$ cannot be explained by the $T=1$ 
pairing Hamiltonian (\ref{eq:2}). 

In a previous paper \cite{kaneko}, it has been shown that the $T=0$ matrix elements 
of the monopole field $V_{m}^{T}(a,b)$ are significantly larger than the $T=1$ ones, 
and are very important in determining the double 
differences of binding energies, where $a, b$ are the single particle orbitals. 
We can see that the matrix elements are quite large for isoscalar components but 
small for isovector components. In the USD interaction, the monopole matrix elements 
with $T=0$ have values around -3 MeV and are strongly attractive. 
If we assume that the $T=0$ monopole matrix 
elements are equal and independent of angular momentum $J$ and the single particle orbitals, 
$V_m^{T=0}$ is reduced to the $J$-independent isoscalar {\it p-n} pairing 
interaction. 
Neglecting $T=1$ monopole components, let us add the $J$-independent $T=0$ $pn$ pairing 
interaction \cite{kaneko,hasegawa} to the pairing Hamiltonian (\ref{eq:2}) 
\begin{eqnarray}
H   & = & H_{0} + H_{P} + H_{\pi\nu}^{\tau=0} \nonumber \\
    & = & H_{0} + H_{P} 
      - k^{0}\sum_{a\geq b}\sum_{J,M}A_{JM,00}^{\dagger}(ab)A_{JM,00}(ab),
\label{eq:4}
\end{eqnarray}
where $A_{JM,00}^{\dagger}(ab)$ is the pair operator with spin $J$ and isospin $T=0$. 
The $T=1$ pairing interaction does not contribute to the double difference of binding 
energies $\Delta_{pn}^{T=0}$, and $\Delta_{pn}^{T=0}\approx k^{0}/2$. 
Then, the $T=0$ pairing force strength $k^{0}=244.5(1-1.67A^{-1/3})/A$ is 
chosen so as to fit the $T=0$ $pn$ pairing gap as seen in Fig. 2. 
 The isovector monopole components in USD are small, except for 
 $V_{m}^{T=1}(s_{1/2},s_{1/2})$. 
 The deviations from the curve $k^{0}/2$ for $^{30}$P and $^{34}$Cl in Fig. 2 would be 
 attributed to the large value of isovector component $V_{m}^{T=1}(s_{1/2},s_{1/2})$. 
 We recently introduced \cite{hasegawa} monopole corrections to improve
  the energy levels of $^{48}$Ca, etc. 
 In this paper, we ignore these correction terms. 

If we assume degenerate single-particle energies $\varepsilon_{a}=0.0$, the above Hamiltonian 
has SO(5) symmetry \cite{Hecht} and the eigenenergy is assigned by the valence nucleon number $n$ and 
isospin $T$ \cite{kaneko}, 
 \begin{eqnarray}
   \langle H_{P_0}+H^{\tau =0}_{\pi \nu} \rangle_{SO(5)}
            = -\frac{1}{2}Gn\left( \Omega - \frac{n-6}{4} \right) \nonumber \\
              -{1 \over 2} k^0 {{n} \over 2}\left( {{n} \over 2} +1 \right)
              + \frac{1}{2}(G+k^{0})T(T+1), \label{eq:5} 
\end{eqnarray}
where $\Omega=\sum_{\alpha}$ is the degeneracy of shell orbits. 
Note that the above equation includes the so-called symmetry energy term with coefficient 
$a(A)/A=(G+k^{0})/2$. 
The parameters $G$ and $k^{0}$ used above give just the empirical symmetry energy formula 
$a(A)=134.4(1-1.52A^{-1/3})$ determined by Duflo and Zuker \cite{Duflo}. 
\begin{figure}[t]
\includegraphics[width=9cm,height=7cm]{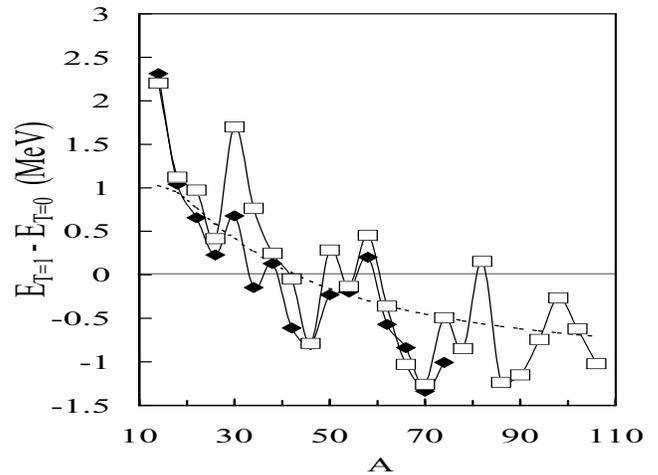}
  \caption{The energy difference between the $T=0$ and $T=1$ states in odd-odd $N=Z$ nuclei. 
  The experimental values of the differences are denoted by solid diamonds. 
  The open squares present the values estimated from the experimental odd-even mass 
  differences in Fig. 1 and the $T=0$ pairing force strength $k^{0}$. 
  The dashed line is $k^{0}-10.4A^{-1/3}$.}
  \label{fig3}
\end{figure}

We next consider energy difference between the lowest $T=0$ and $T=1$ states in odd-odd 
$N=Z$ nuclei. 
Odd-odd $N=Z$ nuclei with $A < 40$ have the ground states with $T=0, J > 0$ except for 
$^{34}$Cl, while the ground states of odd-odd $N=Z$ nuclei with $40 < A < 74$ are 
$T=1$ and $J=0$ except for $^{58}$Cu. 
Several authors discussed that this degeneracy is attributed to the delicate 
balance between the symmetry energy and pairing correlations, 
and that the energy difference between $T=1$ and $T=0$ states is well 
reproduced by $E_{T=1}-E_{T=0}=2a(A)/A-2\Delta_{T=1}$ using the value $\sim 75$ for 
$a(A)$ and the pairing gap $\Delta_{T=1}=12A^{-1/2}$. 
However, if we substitute the odd-even mass difference $\Delta^{(3)}_{n}(Z={\rm even},Z+1)$ 
for $\Delta_{T=1}$, the energy difference $E_{T=1}-E_{T=0}$ becomes larger than the 
experimental value. 
The energy difference can be regarded as a measure of competition 
between the $T=0$ and $T=1$ pairing correlations as seen from the following identity, 
 \begin{eqnarray}
E_{T=1}-E_{T=0} & = & 2(\Delta_{pn}^{T=0}-\Delta_{pn}^{T=1}). 
\end{eqnarray}
The relationships $\Delta_{pn}^{T=0}\approx k^{0}/2$ and 
$\Delta_{pn}^{T=1}\approx\Delta^{(3)}_{n}$ offer an alternative relation 
$E_{T=1}-E_{T=0}\approx k^{0}-2\Delta^{(3)}_{n}$ for the energy 
difference except for $^{30}$P and $^{34}$Cl. 
If we adopt the parameter $k^{0}=244.5(1-1.67A^{-1/3})/A$ and the average value of 
pairing gap $5.18A^{-1/3}$ for $\Delta^{(3)}_{n}$, 
we get the dashed curve in Fig.3, which displays well the trend of the experimental values 
of energy difference $E_{T=1}-E_{T=0}$. 
Adopting the experimental odd-even mass differences for $\Delta^{(3)}_{n}$ and 
$k^{0}=244.5(1-1.67A^{-1/3})/A$, we obtain the energy difference $E_{T=1}-E_{T=0}$ denoted 
by the open squares. 
These values nicely reproduce the experimental values except for $^{30}$P and $^{34}$Cl 
as shown in Fig. 3. 
The disagreements in $^{30}$P and $^{34}$Cl are attributed to the large deviations of $T=0$ 
pairing gap from the curve $k^{0}/2$ due to the neglect of the shell effects in Fig. 2. 

Moreover, we calculated the $T=0$ and $T=1$ energy differences for odd-odd $N=Z$ nuclei with 
$A \geq 78$, although there are no experimental data of the energy difference. 
The calculation predicts that $^{82}$Nb has possibly the ground state with $T=0$, while the other 
odd-odd $N=Z$ nuclei have the $T=1$ ground state. We call this isospin inversion hereafter. 
It is well known that a similar isospin inversion occurs at $^{58}$Cu. 
The isospin inversion is due to characteristic situation, where the Fermi energy lies 
between large spin and small spin orbits with large energy gap i.e., $1f_{7/2}$ and $2p_{3/2}$ 
for $^{58}$Cu, and $1g_{9/2}$ and $2p_{1/2}$ for $^{82}$Nb. 
In these cases, the $T=1$ pairing gap is quite small as seen in Fig. 1, and energy difference 
becomes large from the simple relation 
$E_{T=1}-E_{T=0}\approx k^{0}-2\Delta^{(3)}_{n}(Z={\rm even},Z+1)$. 
\begin{figure}[t]
\includegraphics[width=8cm,height=8cm]{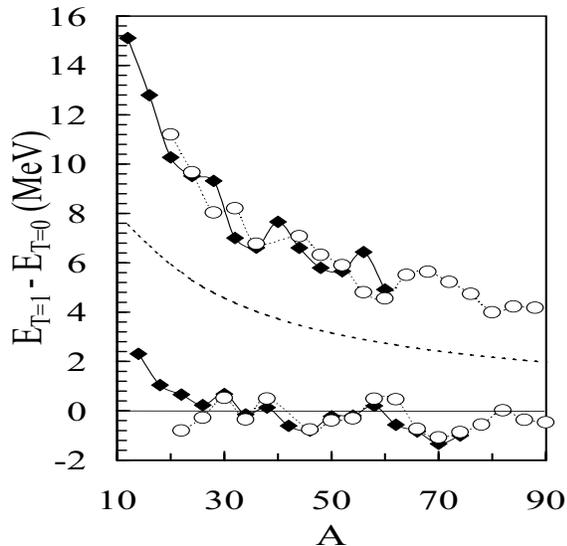}
  \caption{The calculated energy differences between the lowest $T=0$ and $T=1$ states 
  in even-even (upper plots) and odd-odd (lower plots) $N=Z$ nuclei. 
  The solid diamonds are the same as Fig. 3. 
  The open circles denote the energy differences obtained by the 
  BCS calculations. The dashed curve is $2a(A)/A$.}
  \label{fig4}
\end{figure}

Figure 4 shows the calculated energy differences $E_{T=1}-E_{T=0}$ in odd-odd and even-even 
$N=Z$ nuclei. 
The energy differences in the BCS approximations are calculated by $2a(A)/A+\Delta_{BCS}$ 
for even-even $N=Z$ nuclei and by $k^{0}-2\Delta_{BCS}$ for odd-odd $N=Z$ nuclei 
where $a(A)$ is the empirical symmetry energy coefficient and $\Delta_{BCS}$ is the BCS 
pairing gap. The BCS calculations well reproduce the experimental 
values of energy differences, except for odd-odd $N=Z$ nuclei with $A < 40$. 
The BCS calculations show that the $T=0$ and $T=1$ states in $^{82}$Nb are almost degenerate, 
while the ground states of adjacent odd-odd $N=Z$ nuclei have isospin $T=1$. 

In conclusion, we investigated the $T=0$ and $T=1$ pairing correlations in $N=Z$ nuclei. 
The $T=1$ pairing correlations in $N=Z$ nuclei are extracted from the odd-even mass differences 
of the neighboring odd-mass nuclei, which can be fitted by the curve $5.18A^{-1/3}$. 
The $pn$ pairing correlations are estimated from the double difference of binding energies. 
The $T=1$ $pn$ pairing gap is the same as the $nn$ pairing gap. 
The indicator $\Delta_{pn}^{T=0}$ presents the magnitude of $T=0$ $pn$ pairing correlations. 
The energy differences between the $T=0$ and $T=1$ states are well described by 
the $T=1$ and $T=0$ pairing model. 
In odd-odd $N=Z$ nuclei, the $T=1$ pairing correlations compete with the $T=0$ 
pairing correlations, and  the degeneracy of the $T=0$ and $T=1$ states occurs. 
The empirical values and BCS results reproduced 
the energy difference. In particular, our results predict that odd-odd $N=Z$ nucleus $^{82}$Nb 
has the $T=0$ ground state or the $T=0$ and $T=1$ states are almost degenerate. 
The odd-even mass differences for even-even $N=Z$ nuclei are extremely 
larger than those of the neighboring even-even $N\neq Z$ nuclei. 
It would be affected by strong $pn$ correlations. 
Further studies in this direction are in progress.



\end{document}